\documentclass[10pt,twocolumn]{article} 
\usepackage{amsmath,amsfonts}
\usepackage{algorithmic}
\usepackage{algorithm}
\usepackage{array}
\usepackage[caption=false,font=normalsize,labelfont=sf,textfont=sf]{subfig}
\usepackage{textcomp}
\usepackage{stfloats}
\usepackage{url}
\usepackage{verbatim}
\usepackage{graphicx}
\usepackage{cite}
\usepackage[colorlinks=true,linkcolor=black,anchorcolor=black,citecolor=black,filecolor=black,menucolor=black,runcolor=black,urlcolor=black]{hyperref}

\begin{document}

\title{Correction to Friis Noise Factors}

\author{Ankitha E. Bangera \\
\\
\it{Department of Electrical Engineering,} \\
\it{Indian Institute of Technology Bombay,} \\
\it{Mumbai$-$400076, India} \\
E-mail: ankitha\_bangera@iitb.ac.in; ankitha.bangera@iitb.ac.in}
\date{}

\maketitle
\thispagestyle{empty}

\begin{abstract}
The signal-to-noise ratio of a multistage cascade network is often estimated using Friis formulas for noise factors (or the noise figures in decibel). In this letter, the correct formulas to calculate the stage-wise noise factors and the total noise factor in terms of the stage-wise noise factors of an $n$-stage cascade network are derived. A comparison of our derived formulas for noise factors with Friis's formulas is presented. Contrary to Friis's total noise factor in terms of the stage-wise noise factors, we define the actual total noise factor of the $n$-stage cascade network as the product of its stage-wise noise factors.  
\end{abstract}

\section{Introduction}
Cascade systems are widely used in various electrical and electronics engineering domains such as telecommunications and signal processing\cite{ref1,ref2,ref3,ref4,ref5,ref6,ref7}, circuits\cite{ref8,ref9,ref10,ref11,ref12,ref13,ref14,ref15,ref16}, networks and systems\cite{ref17,ref18,ref19,ref20}, solid-state devices\cite{ref21,ref22,ref23,ref24,ref25,ref26,ref27,ref28,ref29}, and so on. To extract the actual signal component at the output of these systems, it is critical to calculate its total noise factor ($F_T$). Friis's formulas for noise are most commonly used to calculate the total noise factor of $n$-stage cascade networks\cite{ref1,ref2,ref3,ref4,ref5,ref6,ref7,ref8,ref9,ref10,ref11,ref12,ref13,ref14,ref15,ref16,ref17,ref18,ref19,ref20,ref21,ref22,ref23,ref24,ref25,ref26,ref27,ref28,ref29}. 

In this letter, we briefly discuss the existing theory to calculate the noise factors of a cascade structure and the well-known Friis formulas for the stage-wise and the total noise factors of a cascade structure. We then derive the correct formulas for the stage-wise noise factor and the total noise factor in terms of the stage-wise noise factors of $n$-stage cascade networks. Our derived formulas are then compared with Friis's formulas, with a discussion on the correction required to Friis noise factors.  

\section{Theory and Discussion}
\begin{figure}[!t]
\centering
\includegraphics[width=3.3in]{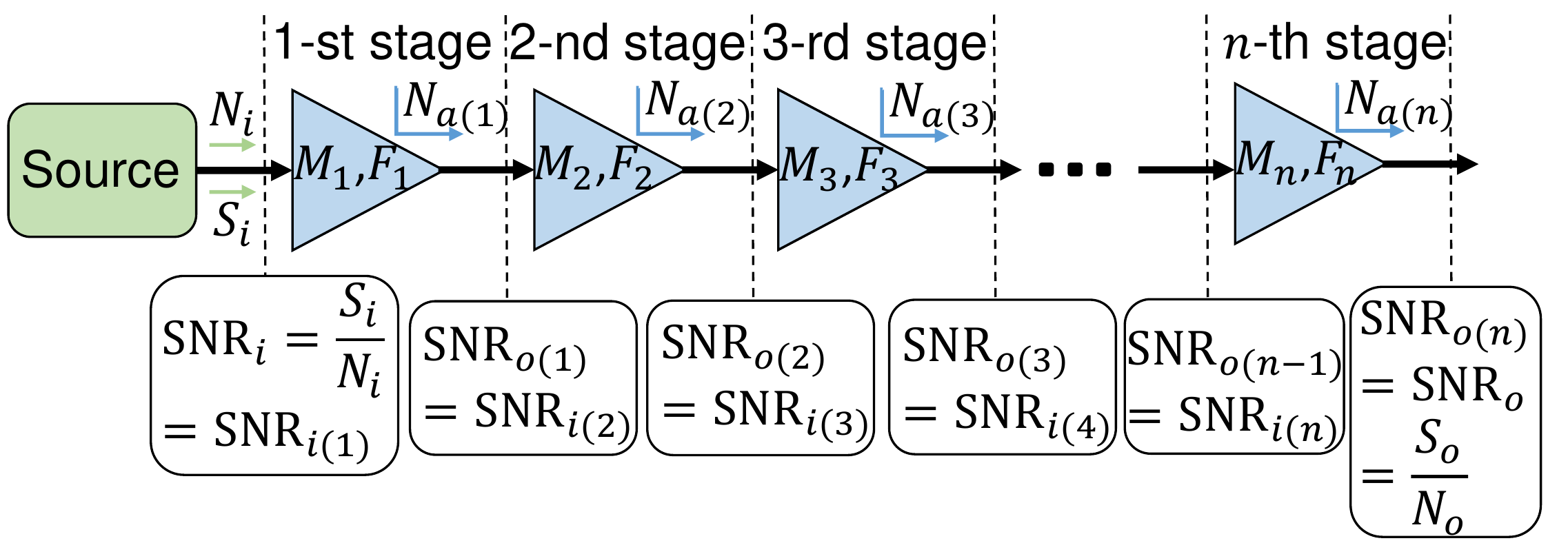}
\caption{Block diagram of an $n$-stage cascade network.}
\label{fig_1}
\end{figure}

Fig.~\ref{fig_1} shows the block diagram of an $n$-stage cascade network. From the block diagram, $S_i$ is the input signal to the network from a source, $N_i$ is the noise from the input source, $M_x$ is the gain of the $x$-th stage, $F_x$ is the noise factor at the $x$-th stage, $N_{a(x)}$ is the added noise at the output of the $x$-th stage, SNR$_{i(x)}$ is the input signal-to-noise ratio (SNR) of the $x$-th stage, SNR$_{o(x)}$ is the output SNR of the $x$-th stage, SNR$_i$ is the input SNR of the network, SNR$_o$ is the output SNR of the network, $S_o$ is the output signal of the network, and $N_o$ is the output noise of the network.

\subsection{Existing Noise Theory and Friis Noise Factors}
The total noise factor of the $n$-stage cascade network ($F_T$) is defined as the ratio of input SNR to output SNR\cite{ref1,ref15,ref30}. Solving for $F_T$ of the $n$-stage cascade network shown in Fig. 1, we get $F_T$ as given by equation~\eqref{eqn_1}.

\begin{equation}
\label{eqn_1}
\begin{aligned}
F_T &= \frac{\text{SNR}_i}{\text{SNR}_o} = \frac{\left(\frac{S_i}{Ni}\right)}{\left(\frac{S_o}{N_o}\right)} \\
&= \frac{\left(\frac{S_i}{N_i}\right)}{\left(\frac{S_i\{\prod_{x=1}^{n}M_x\}}{N_i\{\prod_{x=1}^{n}M_x\}+\sum_{x=1}^{n}N_{a(x)}\{\prod_{y=(x+1)}^{n}M_y\}}\right)}\\
&= 1+\frac{N_{a(1)}}{N_iM_1}+\frac{N_{a(2)}}{N_iM_1M_2}+...+\frac{N_{a(n)}}{N_iM_1M_2...M_n} 
\end{aligned}
\end{equation} 

According to Friis\cite{ref1}, the noise factor at the $x$-th stage of a cascade network is defined by,

\begin{equation}
\label{eqn_2}
F_x^{\text{Friis}} = 1+\frac{N_{a(x)}}{N_iM_x}.
\end{equation}

Substituting equation~\eqref{eqn_2} in~\eqref{eqn_1}, we get, 

\begin{equation}
\label{eqn_3}
\begin{aligned}
F_T =& \underbrace{1+\frac{N_{a(1)}}{N_iM_1}}_{\text{$=F_1^{\text{Friis}}$}}+\underbrace{\frac{N_{a(2)}}{N_iM_1M_2}}_{\text{$=\left(F_2^{\text{Friis}}-1\middle)\middle/\middle(M_1\right)$}}\\
&+...+\underbrace{\frac{N_{a(n)}}{N_iM_1M_2...M_n}}_{\text{$=\left(F_n^{\text{Friis}}-1\middle)\middle/\middle(\prod_{x=1}^{(n-1)}M_x\right)$}}. 
\end{aligned}
\end{equation}

Therefore, Friis defines the total noise factor ($F_T^{\text{Friis}}$) in terms of the stage-wise noise factors ($F_x^{\text{Friis}}$) of an $n$-stage cascade network as\cite{ref1,ref4,ref23}, 

\begin{equation}
\label{eqn_4}
F_T^{\text{Friis}} = F_1^{\text{Friis}}+\sum_{x=2}^{n}\left(\frac{F_x^{\text{Friis}}-1}{\prod_{y=1}^{(x-1)}M_y}\right).
\end{equation}

\subsection{Correct Formulas for Noise Factors} 
For cascade networks, the output of the previous stage will be the input to the next stage. Therefore, the stage-wise noise factor must be equal to the ratio of SNR at the input of the stage to SNR at the output of the corresponding stage\cite{ref1,ref15,ref30}. From the block diagram shown in Fig. 1, the noise factor at the $x$-th stage is,

\begin{equation}
\label{eqn_5}
F_x = \frac{\text{SNR}_{i(x)}}{\text{SNR}_{o(x)}} = \frac{\text{SNR}_{o(x-1)}}{\text{SNR}_{o(x)}}.
\end{equation}

Solving equation~\eqref{eqn_5} for the 1-st stage of the network, we obtain the 1-st stage noise factor ($F_1$) same as that of Friis's 1-st stage noise factor ($F_1^{\text{Friis}}$). However, for higher stages ($x\ge2$), the $x$-th stage noise factor will not be equal to the corresponding Friis's stage-wise noise factor. For illustration, considering the noise factor at the 2-nd stage of the network,

\begin{equation}
\label{eqn_6}
\begin{aligned}
F_2 &= \frac{\text{SNR}_{i(2)}}{\text{SNR}_{o(2)}} = \frac{\text{SNR}_{o(1)}}{\text{SNR}_{o(2)}}\\
&= \frac{\left(\frac{S_iM_1}{N_iM_1+N_{a(1)}}\right)}{\left(\frac{S_iM_1M_2}{N_iM_1M_2+N_{a(1)}M_2+N_{a(2)}}\right)}\\
&= \frac{N_iM_1M_2+N_{a(1)}M_2+N_{a(2)}}{\left(N_iM_1+N_{a(1)}\right)M_2}=\frac{N_{o(2)}}{N_{i(2)}M_2}.
\end{aligned}
\end{equation}

From equation~\eqref{eqn_6}, the stage-wise noise factor may also be defined as the ratio of the total noise at the output of the stage ($N_{o(x)}$) to the total noise at its input ($N_{i(x)}$) multiplied by the stage gain ($M_x$). This definition does not agree with Friis's stage-wise noise factors. A comparison of the stage-wise noise factor for the 2-nd stage of the cascade network is shown in equation~\eqref{eqn_7}. 

\begin{equation}
\label{eqn_7}
\begin{aligned}
F_2 &= 1+\frac{N_{a(2)}}{\left(N_iM_1+N_{a(1)}\right)M_2}=\frac{N_{o(2)}}{N_{i(2)}M_2}=F_2^{\text{Cor}}\\
&\neq\left(F_2^{\text{Friis}}=1+\frac{N_{a(2)}}{N_iM_2}\right)
\end{aligned}
\end{equation}

Therefore, the correct generalized formula for the stage-wise noise factor at the $x$-th stage ($F_x^{\text{Cor}}$) must be,

\begin{equation}
\label{eqn_8}
\begin{aligned}
F_x^{\text{Cor}} &= \frac{N_{o(x)}}{N_{i(x)}M_x} = 1+\frac{N_{a(x)}}{N_{i(x)}M_x}\\
&=1+\frac{N_{a(x)}}{N_i\prod_{j=1}^{x}M_j+\sum_{k=1}^{(x-1)}\{N_{a(k)}\prod_{l=k+1}^{x}M_l\}}.
\end{aligned}
\end{equation}

\begin{figure}[!t]
\centering
\includegraphics[width=2.4in]{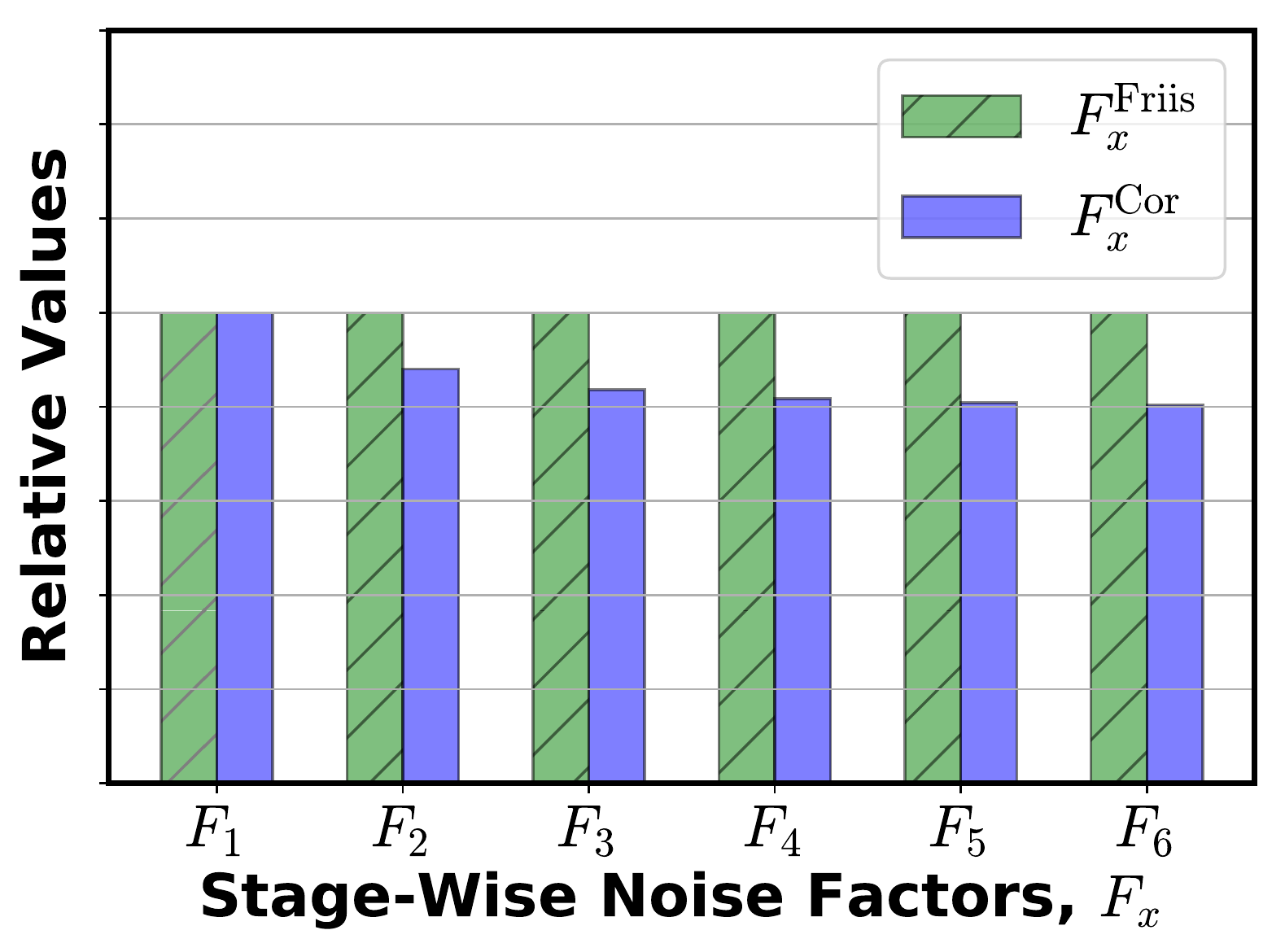}
\caption{A comparison of Friis and our stage-wise noise factors for 1-st to 6-th stages, if there exists a stage-wise added noise that is `greater than zero' and `equal at all the stages'.}
\label{fig_2}
\end{figure}

Comparing equations~\eqref{eqn_2} and~\eqref{eqn_8}, the stage-wise noise factors will be `equal to one' if the stage-wise added noises are `equal to zero'. Thus, if $N_{a(x)}=0$, then, $F_x=F_x^{\text{Cor}}=F_x^{\text{Friis}}=1$. However, if there exists a stage-wise added noise that is `greater than zero' and `equal at all the stages', then, a bar chart comparing the relative values of the stage-wise noise factors calculated using Friis formula (equation~\eqref{eqn_2}) and our formula (equation~\eqref{eqn_8}) for up to the 6-th stage is shown in fig.~\ref{fig_2}. Here, firstly it is observed that the 1-st stage noise factors calculated using Friis and our formulas are equal. Whereas, for higher stages, the noise factors calculated using Friis formula are greater than the corresponding stage-wise noise factors calculated using our formula. Secondly, the Friis stage-wise noise factor values remain the same for all stages if the stage-wise added noises are equal. However, our formula suggests that if all the stage-wise added noises are equal and greater than zero, then the stage-wise noise factors reduce with the stage number. This is because, as the stage number increases, the total noise at its input also increases. Thus, if $\forall x=\{1,2,...,n\}~\exists N_{a(x)}>0 \ni\{N_{a(1)}=N_{a(2)}=...=N_{a(n)}\}$, then, (i) $F_x^{\text{Friis}}=F_x^{\text{Cor}}$ for $x=1$ and $F_x^{\text{Friis}}>F_x^{\text{Cor}}$ for $x\geq2$; (ii) $F_1^{\text{Friis}}=F_2^{\text{Friis}}=...=F_n^{\text{Friis}}$, whereas, $F_1^{\text{Cor}}>F_2^{\text{Cor}}>...>F_n^{\text{Cor}}$. Therefore, our formula for the stage-wise noise factor is a correction to Friis's formula. 

Moreover, the correct generalized formula for the stage-wise noise factor at the $x$-th stage in terms of the stage-wise noise factors of the previous stages is written as,

\begin{equation}
\label{eqn_9}
F_x^{\text{Cor}} = 1+\frac{N_{a(x)}}{N_i\prod_{j=1}^{x}M_j\prod_{k=1}^{(x-1)}F_k^{\text{Cor}}}.
\end{equation}

Rearranging equation~\eqref{eqn_9} and substituting it in equation~\eqref{eqn_1} we get, 

\begin{equation}
\label{eqn_10}
\begin{aligned}
F_T =& \underbrace{1+\frac{N_{a(1)}}{N_iM_1}}_{\text{$=F_1^{\text{Cor}}$}}+\underbrace{\frac{N_{a(2)}}{N_iM_1M_2}}_{\text{$=\left(F_2^{\text{Cor}}-1\right)F_1^{\text{Cor}}$}}\\
&+...+\underbrace{\frac{N_{a(n)}}{N_iM_1M_2...M_n}}_{\text{$=\left(F_n^{\text{Cor}}-1\right)\prod_{x=1}^{n-1}F_x^{\text{Cor}}$}}. 
\end{aligned}
\end{equation}

Therefore, the correct total noise factor of the $n$-stage cascade network ($F_T^{\text{Cor}}$) in terms of the stage-wise noise factors ($F_x^{\text{Cor}}$) is given by equation~\eqref{eqn_11}, which is not equal to equation~\eqref{eqn_4}. 

\begin{equation}
\label{eqn_11}
F_T^{\text{Cor}} = \prod_{x=1}^{n}F_x^{\text{Cor}}\neq\left(F_T^{\text{Friis}}\right)
\end{equation}

From equation~\eqref{eqn_11}, the actual total noise factor of the $n$-stage cascade network is defined as the product of the stage-wise noise factors.

\section{Conclusion} 
We conclude that our derived formulas for the stage-wise noise factor for stages $x\ge2$ and the total noise factor in terms of the stage-wise noise factors of the $n$-stage cascade network are a correction to Friis's formulas.

\section*{Acknowledgments}
A.E.B. would like to thank the Indian Institute of Technology Bombay for their support.

\bibliographystyle{unsrt}
\bibliography{BibRef}

\end{document}